\def\ifmath#1{\relax\ifmmode #1\else $#1$\fi}%

\def\be{\begin{equation}}
\def\ee{\end{equation}}
\def\bea{\begin{eqnarray}}
\def\eea{\end{eqnarray}}
\def\oJ{$\overline{J}$}

\documentclass{ws-p8-50x6-00}

\begin{document}

\title{BUCKYBALLS OF QCD: GLUON JUNCTION NETWORKS}

\author{T. Cs\"{o}rg\H{o}}

\address{MTA KFKI RMKI, H-1525 Budapest 114, POB 49, Hungary\\
        E-mail: csorgo@sunserv.kfki.hu}

\author{M. Gyulassy}

\address{Department of Physics, Columbia University\\
        538 W 120th Street, New York, NY 10027, USA\\
        E-mail: gyulassy@mail-cunuke.phys.columbia.edu}

\author{D. Kharzeev}

\address{Department of Physics, Brookhaven National Laboratory\\ 
Upton, New York 11973 - 5000, USA\\
        E-mail: kharzeev@bnl.gov}


\maketitle

\abstracts{We propose that femto scale 
QCD analogs of Fullerene type excitations 
        may exist with ``magic numbers'' 8,24,48, and 120. 
These ``J-balls'' are the most symmetrical 
closed  networks of gluonic baryon junctions and anti-junctions 
with zero net baryon number. 
CP odd J-Moebii and other femto-tubes and tori gluonic structures
 may also exist.
We estimate the relative binding energies
        and discuss possible  experimental signatures of the formation
        of these new topological gluonic states.  }

\section{Introduction}

In QCD, baryons are made of quarks. However, to construct a 
gauge-invariant wave function of the baryon, one has to introduce also an 
interesting configuration of gauge fields -- the ``baryon junction'' \cite{veneziano}. 
The junction is built out of 
Wilson lines or strings, $U(x,\bar{x})
= P \exp \left (ig \int_{\bar{x}}^{x} dx^{\mu} A_{\mu} \right ) 
$, with gauge field $A$. It ensures that the nonlocal baryon wave function, 
with quarks at $x_i$ and a junction at $x$ is invariant under  
gauge transformations. It has a color tensor structure
$J_{j_1,j_2,j_3}(x;x_1,x_2,x_3)= \epsilon_{k_1k_2k_3}
U_{j_1k_1}(x,x_1) U_{j_2k_2}(x,x_2)
U_{j_3k_3}(x,x_3)$. A color neutral baryon with quark flavors 
 (1,2,3) is then constructed as the contraction
$B_{123}(x)=q_1(x_1)q_2(x_2)q_3(x_3)J(x;x_1,x_2,x_3)$. 

It is interesting that junctions in $SU(N)$ gauge theories call 
for the existence of a new family of glueball--like bound states 
in the confined phase, with masses scaling with the number of 
colors, $M \sim N$. (Note that ``ordinary'' glueballs, 
described by the closed strings, have masses $M \sim O(1)$.)     
In QCD, the simplest bound state of 
this kind is a quarkless ``baryonium'' 
J-ball, 
formed by connecting a $J(x;...)$ with an anti-$J(\bar{x};...)$
\cite{veneziano}:
$$ 
M^J_0(x,\bar{x})  = Tr(J(x)J^\dagger(\bar{x}))=
 \epsilon^{j_1j_2j_3} U_{j_1k_1}(x,\bar{x}) U_{j_2k_2}(x,\bar{x})
U_{j_3k_3}(x,\bar{x})\epsilon_{k_1,k_2,k_3}
$$ 
While this state is constructed only from the gluon field, and thus has 
a zero net baryon number, it has a large coupling to the baryon wave functions, 
and couples to the $B-\bar{B}$ state. 

In high--energy collisions, the valence quarks carry a large fraction of the 
incident baryon's momentum, and thus always hadronize in the fragmentation regions. 
This, undoubtedly  correct, statement is at the origin of the widely accepted 
belief that the central rapidity region of high energy proton and nuclear collisions 
should be ``baryon--free'', i.e. should have a zero net baryon number.   
The existence of baryon junctions invalidates this naive picture, since gluonic 
junctions on the average carry only small fraction of the baryon's momentum, 
and thus can be easily transferred to the central rapidity region, where 
the string break-up dresses them by quarks. In other words, this picture  
 implies~\cite{dima96} that the  
        baryon number effectively resides  
        in purely  non-perturbative (topological) configuration of
        gluon fields, rather than in the valence quarks.
        In heavy ion physics, the interest in these
        configurations stems from the possibility 
        that they lead to  substantial 
        baryon asymmetry in the mid-rapidity region 
        of ultra-relativistic heavy ion collisions\cite{dima96}, 
        leading also to 
        new diquark breaking mechanisms \cite{kop89,ck} which 
        could transfer the baryon number over many units of rapidity.  
        These configurations provide a novel mechanism for strangeness
        enhancement~\cite{vance-jb} and  for production of hyperon pairs, too. 
        The resulting  junction model of baryons can also be 
        formulated in a Poincare invariant dynamical model for a baryon,
        where three valence quarks and a junction that represents
        sea quarks and all gluonic degrees of freedom interact as
        classical particles via a quasi potential. In spite of its simplicity,
        this junction model is able to describe some features of the baryon
        (binding energy, structure functions) surprisingly well~\cite{noack}.
        In addition, preliminary RHIC data\cite{nuqm01} 
        were consistent with the high baryon stopping power predicted by
        $M_J^0$ exchange~\cite{vance-jb,dima96}.

        In this study we explore the combinatorial consequences of 
        the existence of networks or closed cages
        involving junction and anti-junction type of configurations
        in QCD (and non-Abelian gauge theories in general).
        In particular, we argue that higher excitations of the QCD vacuum
        can be constructed, that are characterized by remarkable geometrical
       and topological  structures. 
        Certain ``magic numbers" exist with high symmetry similar 
        to Fullerene structures made of carbon, discovered 
        in ref.~\cite{cballs}.  However, in the QCD case the 
        three valence connections are replaced by Wilson lines  
        as in Eq. (1), representing the flow of color flux 
        between adjacent junctions and anti-junctions.
                
        Although the building blocks and the QCD Lagrangian are CP even, 
        states can be constructed that are CP odd.
        We characterize the topological structure of these novel 
        structures in QCD. 

\section{Fullerenes}

        New forms of the element carbon - called fullerenes-
        in which the atoms are arranged in closed shells were
        discovered in 1985 by R. F. Curl, H. W. Kroto and 
        R. E. Smalley~\cite{cballs}. The number of carbon atoms in the shell
        can vary and for this reason numerous new carbon structures
        have become known. 

        Fullerenes are formed when vaporized carbon condenses in an
        atmosphere of inert gas. The gaseous carbon is obtained e.g.
        by directing an intense pulse of laser light at 
        a carbon surface. The released carbon atoms are mixed
        with a stream of helium gas and combine to form
        clusters of some few up to hundreds of atoms.
        The gas is then led into a vacuum chamber where it
        expands and is cools. The carbon clusters can then
        be analyzed with mass spectroscopy.
        
        Curl, Kroto and Smalley performed such an experiment
        and were able in particular to produce clusters with 
        60 and 70 carbon atoms. They found high stability 
        in $C_{60}$ which suggested a molecular structure of
        great symmetry. It was suggested that $C_{60}$
        could be a ``truncated icosahedron cage", a polyhedron
        with 20 hexagonal surfaces and 12 pentagonal
        (5 angled) surfaces. The pattern of European 
        football has exactly this structure, as does the
        geodetic dome designed by the American architect
        R. Buckminster Fuller for the 1967 Montreal World
        Exhibition. The researchers named the
        newly discovered structure {\it ``Buckminsterfullerene"}
        after him, subsequently the nick-named as 
        ``Buckyball".

        This discovery opened the way to a completely
        new branch of chemistry creating new forms of 
        matter with unusual properties and to the
        1996 Nobel prize in chemistry to Curl, Kroto and
        Smalley, see ref.~\cite{nobel_c} for further details.

\section{QCD Buckyballs }
        The QCD junctions and anti-junctions can be combined in higher
        excited states  which can be called as QCD fullerenes (or Buckyballs), 
        that are similar to the  
        nano-structures formed in carbon chemistry, as summarized
        in the previous section.
        The QCD fullerenes (or junction balls) 
        can be characterized by the topology
        of their structure. Due to the existence of junctions (J) and
        anti-junctions (\oJ), the QCD fullerenes may have only even
        number of vertices  along any faces,
        hence their geometrical structure will be 
        different from the carbon fullerenes, that contain
        12 pentagonal and a varying number of hexagonal faces.

        Let us recall Euler's theorem that relates the number of faces (F),
        the number of edges (E) and the number of vertices (V) in a
        polyhedron as
\be
        V + F = 2 + E
\ee
        where the number of faces is the sum of the number of squares ($N_4$),
        the 
\begin{figure}[t]
\centerline{\epsfxsize=7pc 
\epsfbox{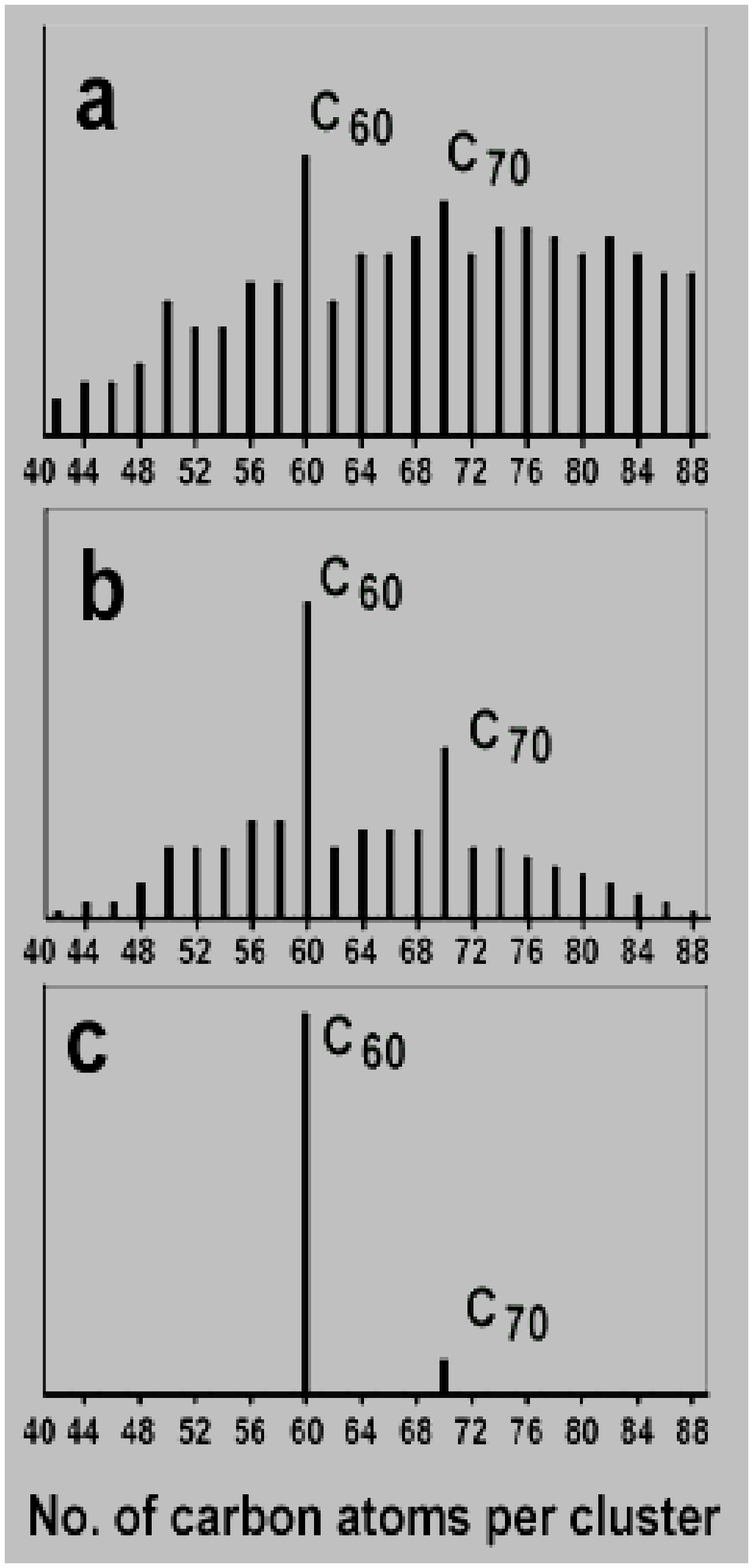}
\epsfxsize=20pc 
\epsfbox{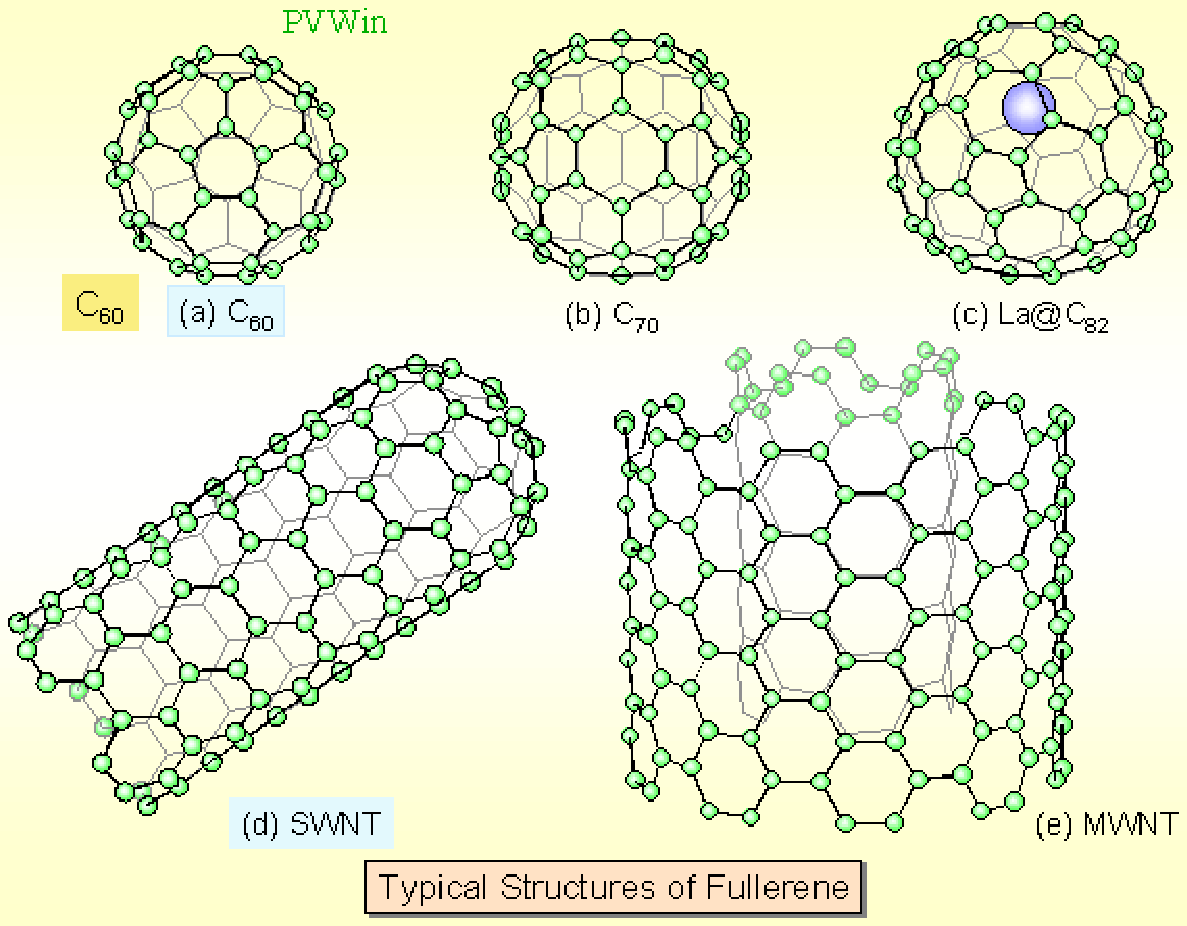}} 
\caption{(Left) Discovery of carbon Buckyballs $C_{60}$, 
$C_{70}$ amidst the ashes of burnt graphite
ignited  by  a laser {\protect\cite{cballs,nobel_c}}. (Right) Cage structure of Buckyballs and tubes
from {\protect\cite{c70_f}}.
\label{f:1}}
\end{figure}
\begin{figure}[h]
\epsfxsize=5pc 
\centerline{
\epsfig{file=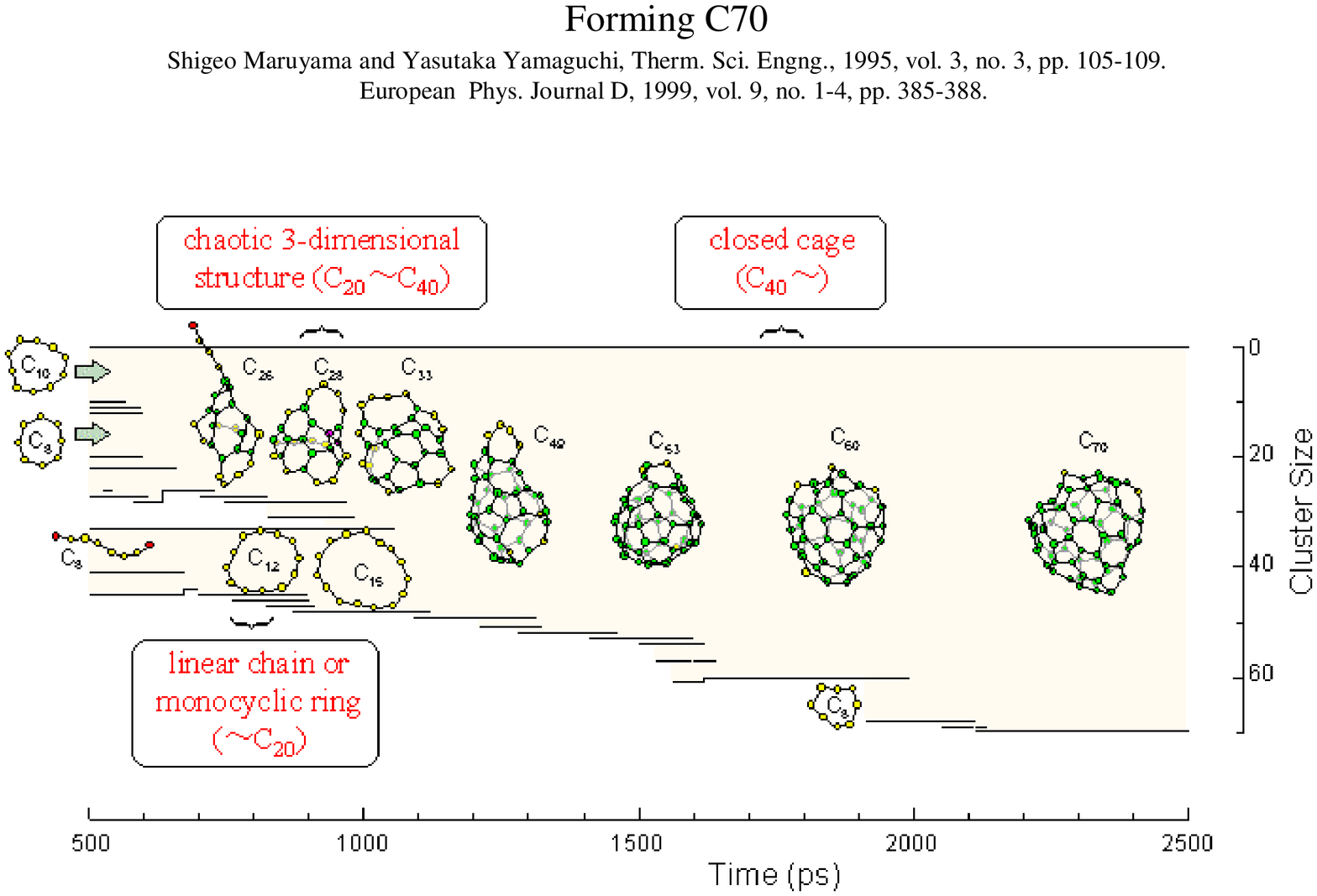,scale=0.5,angle=-90}} 
\vspace{-0.8in}
\caption{Formation of $C_{70}$ through closing carbon chains,
merging them into carbon rings, forming random open cages,
and finally wrapping into  closed carbon cages from {\protect\cite{c70_f}}.
\label{f:2}}
\end{figure}
number of hexagons ($N_6$), the number of octagons ($N_8$), etc,
\be
        F = N_4 + N_6 + N_8 + ...
\ee
        Each edge belongs to two faces. 
        As junctions are color sinks, 
        while anti-junctions are color sources, 
        each junction has to be surrounded 
        by three anti-junctions, and vice-versa, 
        thus each vertex belongs to
        three faces: 
\bea
        E & = & \frac{1}{2} \left( 4 N_4 + 6 N_6 + 8 N_8 + ... \right), \\
        V & = & \frac{1}{3} \left( 4 N_4 + 6 N_6 + 8 N_8 + ... \right).
\eea
        The resulting Diophantic equations are solvable.
        The solution is independent from the number of hexagons,
        $N_6 = any$, the remaining constraint is
\be
        N_4 - \sum_{i=4}^\infty (i - 3) N_{2 i} = 6
\ee
        This implies that if octagons, decagons and other structures with
        higher even number of vertices are neglected, 6 squares and any number
        of hexagons can be utilized to build up Euler polyhedra.
        These correspond to  QCD fullerenes 
        where the junctions and anti-junctions
        alternate on the neighboring vertices.
        Such a variety of structures is similar to the case 
        of carbon nano-structures, where any number of hexagons 
        and exactly 12 pentagons can be combined 
        to create a rich family of carbon nano-structures.

\begin{figure}[h]
\epsfxsize=4pc 
\centerline{\epsfig{file=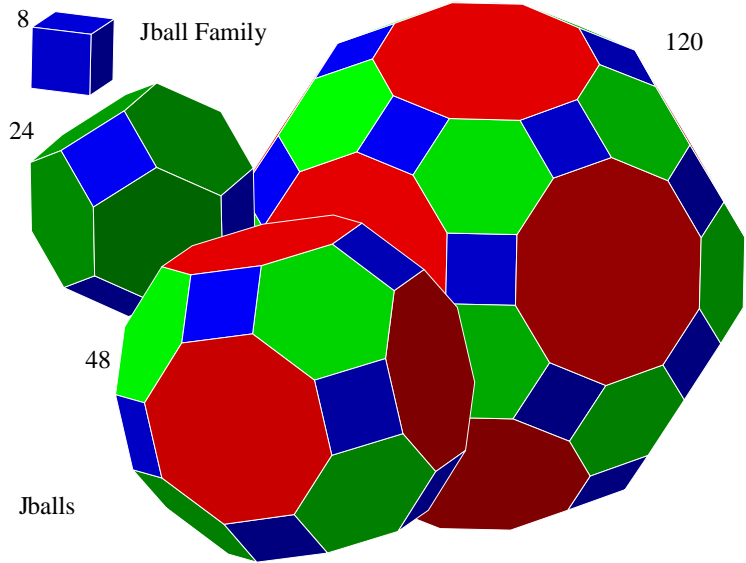,scale=0.4,angle=-90}} 
\caption{The family of QCD fullerenes with magic numbers 8, 24, 48 and
120.  \label{f:3}}
\end{figure}

\section{Magic numbers of QCD fullerenes}

The formation of QCD fullerenes is a complicated dynamical process.
A necessary condition for such process is the (symmetric) presence of
the building blocks, the baryon and anti-baryon junctions.
Thus QCD fullerenes may be excited in processes with vanishing
(or small) net baryon number and high energy density sufficient
to create the anti-baryons in big enough number.  

A similar process is illustrated on Fig. 2 for the case of carbonic
fullerenes~\cite{c70_f}. 
Here we will not speculate further on the dynamics that may
lead to the formation of QCD J-balls. Our aim is simply to try to identify
the most stable configurations  and their ``magic'' (vertex)  numbers.

As in the case of carbon fullerenes, we assume that 
the most symmetric (Archimedean polyhedra) configurations
are  the most stable ones. We estimate the energy of a 
particular configuration and the relative binding energies in the next
paragraph. It is straightforward to solve the Euler equations
for the case of the regular polyhedra. There exists only four 
Archimedian polyhedra consisting of even 
number of junction and anti-junction
pairs. These numbers together with relative size and shape are
shown in Figure 3. Many other, less regular shapes, of course, may
exists, but these are the ones with highest symmetry. 
In $SU(N)$ gauge theories these structures are replaced by 
the most symmetric cages formed with  each vertices connecting
to exactly $N$ edges.
It follows that in $SU(2)$ only ring-like structures can be created.

The simplest model for the relative energy of a J-ball consisting of
$n_V/2 $ junctions and $n_V/2$ anti-junctions making a polyhedron
characterized by $n_E$ edges with lengths $\ell_i$ is given by
\be
E(\ell_i, n_{v,i}; N_V, N_E) = \sum_{i=1}^{N_E} 
\left(\frac{a}{\ell_i} + \kappa \ell_i
\right)+ \gamma \sum_{v}^{N_V} \sum_{i<j} \hat{n}_{v,i}\cdot\hat{n}_{v,j} 
\ee
where $\hat{n}_{v,i=1,2,3}$ are the three unit vectors along the edges 
joining vertex $v$ that define the topology.
 The first term is a ``kinetic'' vertex localization energy.
The second is the confining string tension 
($\kappa\approx 1-\frac{\pi}{3}T^2+...$ GeV/fm) term that vanishes at 
$T_c\approx 150$ MeV.
The last postulated ``strain'' term is analogous to 
the Biot-Savart law in circuits and plays the same role
as bond angle strain in
carbon nanostructures\cite{c70_f}. 

In this model the relative binding energies are determined by the last term,
given by the geometrical structure of the J-ball.
For the $N_V=8$ J-cube $\sum \hat{n}_{v,i}\cdot\hat{n}_{v,j}=0$,
while for $N_V=24,48,120$ , this strain energy is
$-1, -(1+\surd 2)/2=-1.207,-(3+\surd{5})/4=-1.309$ in units of $\gamma$. 
        The absolute minimum
of $(J\bar{J})^{N_V/2}$ Jballs is reached of course for $N_V=\infty$,
corresponding to a hexagonal (graphite-like) flat 2D lattice, as in the
case of carbon fullerenes.

\section{CP Odd and Even J-Ribbons}

Other multi  $(J\bar{J})^N$ configurations can be constructed 
that have definite CP symmetry transformation properties.

In case of carbons, the elementary building block of fullerene
formation is a ring or a closed chain of carbons. In case of QCD,
this is replaced by ribbons that are formed from $(J\bar{J})$ pairs.
Closed ribbons have different CP eigenstates depending on the 
number of $(J\bar{J})$ pairs in them. For even number of pairs,
the ribbon can be closed on itself without any twist, hence
 $(J\bar{J})^{2n}$ states can form CP even J-prisms
as indicated on the left panel of Fig 4.
However, other intriguing possibilities also exist.
In particular,  for odd number of $N=2n+1$    $(J\bar{J})$ pairs 
the QCD ribbon can be closed on itself with a rotation to the left
or with a rotation to the right, forming a Moebius ribbon with 
winding number +1 or -1. These states transform into each other,
a class of CP-odd J-Moebii is obtained. 
Note that for large values of $N$ the ribbon can be twisted not only
once but many times before closing the surface on itself. 
It seems to be rather interesting to explore the CP and topological
properties of these excitations; the closed ribbons can be
characterized with their topological winding number that can
be any integer number $N$,  $N=0$ corresponding to the simplest,
 CP even J-prisms. 
\begin{figure}[t]
\epsfxsize=4pc 
\centerline{\epsfig{file=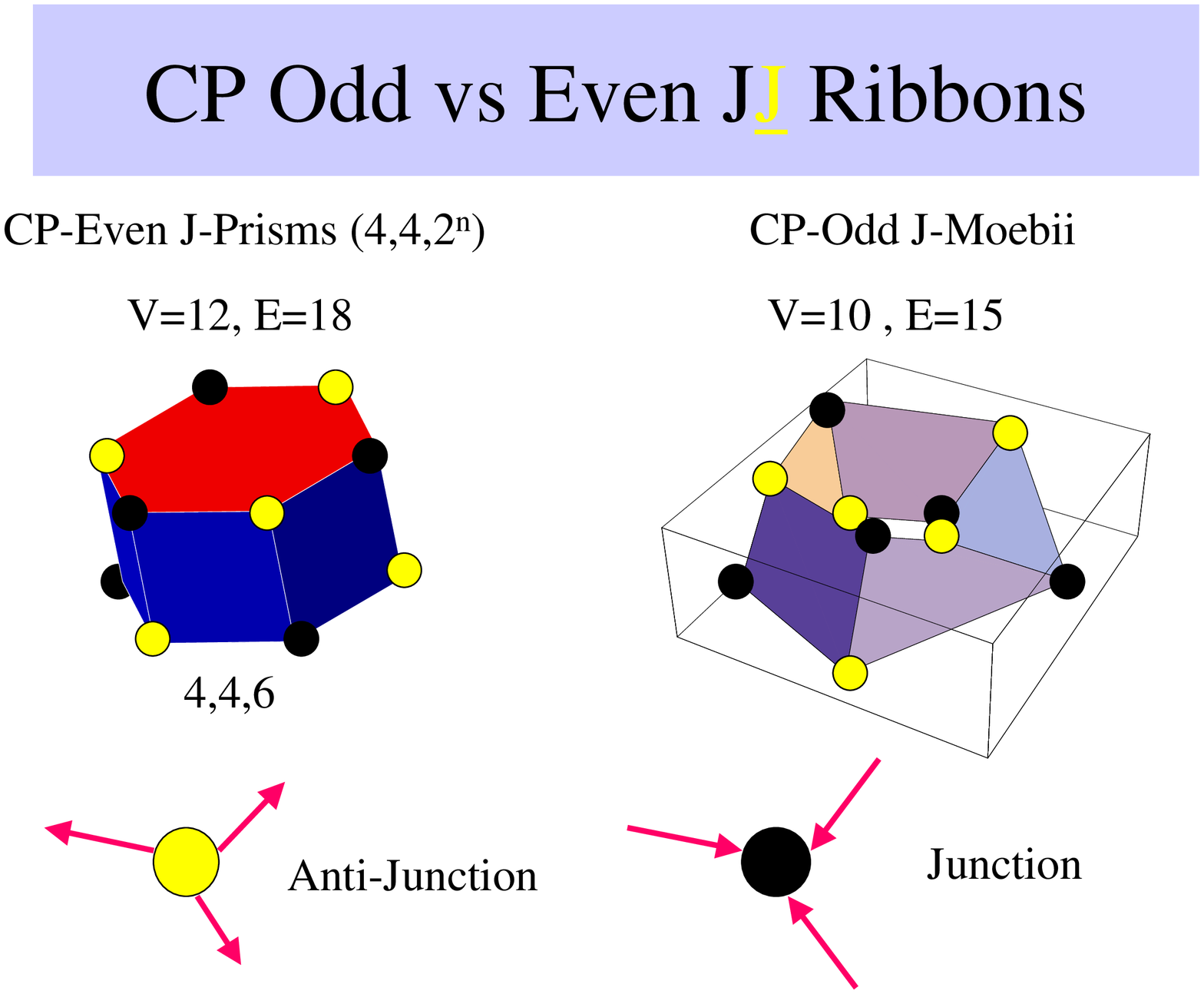,scale=0.5,angle=-90} }
\caption{A CP even J-ribbon and a CP-odd J-Moebius-ring.
\label{f:4}}
\end{figure}
 
The topology of the excitations of QCD may be  very interesting,
because not only ribbons but also tubes can be formed.
These would be analogous to
the carbon nano-tubes displayed
on Fig. 1. The ends can be closed with caps formed by squares,
octagons and decagons, or can be open, closed with the help
of valence quarks. 
Another interesting possibility is to close the J-tube
on itself, creating a toroidal structure.  
Carbon tubes are on the nano meter scale. The QCD excitations exist
on the femtometer scale, hence J-tubes can be considered
as femto-tubes. 

The carbon nano-tubes can be characterized by two integers,
these can be obtained from cutting a planar hexagonal lattice
and using the number of steps in the direction of one of the 
edges in a pre-selected vertex to find the point that is identified
with the same vertex in the planar hexagonal lattice, these
are called the $(n,m)$ nano-tubes.

The QCD femto-tubes have different topological property due to the
existence of junctions and anti-junctions. This implies that on the
planar hexagonal lattice a junction has 6 other anti-junctions in the  
nearest neighbor position, indicating also 2 independent directions.
Hence the QCD femto-tubes can also be characterized by 2 integers,
indicating the number of steps in these directions to find the 
equivalent junctions in the planar hexagonal lattice that are
identified to close the planar structure to a tube. These can
structures be classified as $(n,m)$ QCD femto-tubes.
The femto-tubes can be
closed also by connecting the two ends of a long tube, and these
ends can be rotated before the connection. This gives QCD 
femto-tori that can be characterized by 3 winding numbers,
the $(i,j,k)$ femto-tori.

\section{Estimates and observables}

        We estimate that the structures with the smallest strain
        and the smallest total baryon number
        are the easiest to produce experimentally. This suggests that
        the Archimedian polygons are favored, where each vertex carries
        the same amount of strain. The contribution from the extra
        strains as compared to a planar structure depends on a
        parameter $\gamma$ that is not yet known; but the relative strain
        energy decreases from 
        0 to -1 from the cube to the $B+\overline{B}$ = 24
        J-ball; it decreases further by 20 \% when forming the
        $B+\overline{B}$ = 48 polyhedron. The next Archimedian 
        polyhedra has $B+\overline{B}$ = 120, which is probably
        too large even for RHIC conditions.
         Hence we suggest to search for
         clusters that decays to equal number of baryons and anti-baryons
        with a total number of 8, 24 and 48.

        In addition, we suggest  to search for clusters of equal number
        of baryons and anti-baryons that violate CP symmetry,
        as these clusters also appear in the spectrum of QCD Jballs.
        Some experimental observables that may be used to search
        for the formation of CP violating  
        domains or
         clusters were summarized in the talk of J. Sandweiss~\cite{cp-sum}.

\section{Summary and outlook}

        In summary, fullerene type of pure glue topological configurations
        exist in QCD. These are termed as junction-balls (Jballs) or 
        QCD femto-structures. All of these configurations have
        an equal number of junctions and anti-junctions, but 
        have interesting geometrical and topological properties,
        that can be determined in a straightforward manner.
        Some of the QCD Buckyballs are CP even, others like Moebius
        ribbons are CP odd states. Topological winding 
        numbers can be introduced to characterize these states.
        The QCD femto-ribbons are characterized by 
        a single integer $(n)$, the  femto-tubes 
        by a pair of integers $(n,m)$, while the QCD femto-tori
        by a triplet of integers, $(n,m,l)$.

        We determined that the most symmetric (likely most stable)
        j-ball configurations have the magic number of 
        baryon + anti-baryon number of $B + \overline{B} =$ 8, 24, 48
        and 120. Although these configurations are likely unstable,
        they are expected to be more stable than clusters of baryons
        and anti-baryons with different junction numbers,
        and they may  appear as peaks in the spectrum of
        $(B\bar{B})^n$ clusters with a given total baryon+antibaryon number.
        To create them, high initial energy densities and
        small net initial baryon number densities and large volumes
        are needed.  Such conditions may exist 
        in the mid-rapidity domain of central $Au+Au$ collisions 
        at RHIC or in the mid-rapidity domain of 
        $p+\overline{p}$ collisions at the Tevatron.

        It would certainly be of great theoretical and experimental interest
        if these novel gluonic 
	states were observed in QCD lattice simulations
        or  in baryon-antibaryon clusters in  high energy reactions.

\section*{Acknowledgments}
T. Cs. thanks Columbia University for 
hospitality during several 
visits and to M. Albrow for
inspiring discussions. MG thanks  
the Collegium Budapest and KFKI. 
This research has been supported by 
a Bolyai Fellowship of the Hungarian Academy of Sciences,
by the Hungarian OTKA grants T024094, T026435, T029158, T034269,
by the US-Hungarian Joint Fund MAKA 652/1998, by the NWO-OTKA grant
N025186 and by the US DOE under contracts no. DE-FG-02-93ER-40764,
and DE-AC02-98CH10886.

\end{document}